\documentclass{elsart}
\usepackage{graphics}
\usepackage{graphicx}
\usepackage{epsfig}
\usepackage{amssymb}

\newcommand{\reactiona}{\mbox{$pp\to ppK^+K^-$}}

\begin{document}

\begin{frontmatter}

\title{Coupled-channel effects in the $\mathbf{\reactiona}$ reaction}
\author[Gatchina]{A.~Dzyuba},
\author[ikp]{M.~B\"uscher},
\author[ikp]{M.~Hartmann},
\author[ikp,Tbilisi]{I.~Keshelashvili}\footnote{Current address: Department of Physics and
Astronomy, University of Basel, CH-4056 Basel, Switzerland},
\author[Gatchina]{V.~Koptev},
\author[Osaka]{Y.~Maeda},
\author[ikp,Bonn]{A.~Sibirtsev},
\author[ikp]{H.~Str\"oher},
\author[CW_College]{C.~Wilkin\corauthref{cor1}}
\ead{cw@hep.ucl.ac.uk} \corauth[cor1]{Corresponding author.}
\address[Gatchina]{High Energy Physics Department, Petersburg Nuclear
Physics Institute, 188350 Gatchina, Russia}
\address[ikp]{Institut f\"ur Kernphysik, Forschungszentrum J\"ulich, 52425
  J\"ulich, Germany}
\address[Tbilisi]{High Energy Physics Institute, Tbilisi State
University, 0186 Tbilisi, Georgia}
\address[Osaka]{Research Center for Nuclear Physics, Osaka
University, Ibaraki, Osaka 567-0047, Japan}
\address[Bonn]{Helmholtz-Institut f\"{u}r Strahlen- und Kernphysik (Theorie),
Universit\"{a}t Bonn, D-53115 Bonn, Germany}
\address[CW_College]{Physics and Astronomy Department, UCL, London, WC1E 6BT, UK}
%
%
%
\begin{abstract}
The cross sections for the \reactiona\ reaction were measured at
three beam energies 2.65, 2.70, and 2.83\,GeV at the COSY-ANKE
facility. The shape of the $K^+K^-$ spectrum at low invariant masses
largely reflects the importance of $K\bar{K}$ final state
interactions. It is shown that these data can be understood in terms
of an elastic $K^+K^-$ rescattering plus a contribution coming from
the production of a $K^0\bar{K}^0$ pair followed by a charge-exchange
rescattering. Though the data are not yet sufficient to establish the
size of the cusp at the $K^0\bar{K}^0$ threshold, the low mass
behaviour suggests that isospin-zero production is dominant.
\end{abstract}

\end{frontmatter}
%
%
The COSY-ANKE collaboration has recently published data on the
differential and total cross sections for the \reactiona\ reaction at
three beam energies $T_p=2.65$, 2.70, and 2.83\,GeV, which correspond
to excess energies of $\varepsilon=50.6$, 66.6, and 108.0\,MeV,
respectively~\cite{Maeda08}. The $K^+K^-$ invariant mass $M_{\rm
inv}(KK)$ spectra show a strong signal for the production and decay
of the $\phi$ meson. This sits upon an apparently non-resonant
background. However, the distributions in the $Kp$ and $Kpp$
invariant masses prove that this background is strongly distorted by
a $K^-p$ final state interaction (\emph{fsi})~\cite{Winter}. After
taking this \emph{fsi} into account, as well as the one between the
outgoing protons, most of the distributions are well described by
Monte Carlo simulations.

\begin{figure}[h]
\begin{center}
\includegraphics[clip,width=7cm]{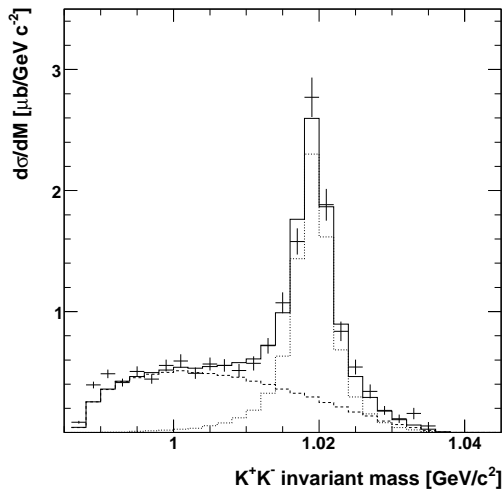}
\caption{Differential cross section for the $pp\to ppK^+K^-$ reaction
at 2.65\,GeV as a function of the $K^+K^-$ invariant mass compared to
Monte Carlo simulations of $\phi$ (dotted) and non-$\phi$ (dashed)
contributions and their sum (solid histogram).} \label{KKmass}
\end{center}
\end{figure}

Nevertheless, as seen in the $K^+K^-$ invariant mass spectrum at
2.65\,GeV shown in Fig.~\ref{KKmass}, the simulation underestimates
the experimental points for $M_{\rm inv}(KK)< 995\,$MeV/c$^2$. Of
itself, this could be dismissed as a fluctuation, though it is
important to realise that the ANKE spectrometer has a very good
acceptance in this region~\cite{ANKE}. Furthermore, a similar
phenomenon was observed for the same mass region in the 2.70 and
2.83\,GeV ANKE data, as well as in those obtained by the DISTO
collaboration at a slightly higher energy~\cite{DISTO}. Moreover, an
analogous effect was also noted for the $pn\to dK^+K^-$ reaction,
where the experimental systematics are rather
different~\cite{Maeda06}.

\begin{figure}[h]
\begin{center}
\includegraphics[clip,width=7cm, angle=0]{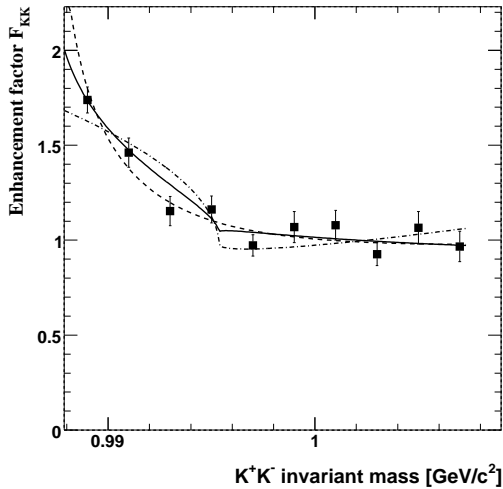}
\caption{Ratio of the $K^+K^-$ invariant mass spectra from the
\reactiona\ reaction to the simulation presented in
Ref.~\cite{Maeda08}. The experimental points correspond to the
weighted average of data taken at 2.65, 2.70, and 2.83\,GeV. The
solid curve is the result of a best fit of Eq.~(\ref{amps}) to these
data, with the parameters being given in Table~\ref{table1}. The
dot-dashed curve is the best fit when the elastic rescattering is
neglected and the dashed when the charge-exchange term is omitted.
\label{Fit}}
\end{center}
\end{figure}

The simulation of the \reactiona\ spectrum of Ref.~\cite{Maeda08},
shown in Fig.~\ref{KKmass} for the 2.65~GeV data, includes only the
final state interactions in the $K^-p$ and $pp$ systems. To
investigate the low $K^+K^-$ mass region in finer detail, we have
taken these results and divided them by the simulation. Although the
resulting error bars are rather large, the ratios at all three
energies are mutually consistent and Fig.~\ref{Fit} shows the
weighted averages of the points at the three energies.

An enhancement at low $K^+K^-$ masses is, of course, to be expected
from an elastic $K^+K^-$ \emph{fsi}, which was not included in the
simulation of the ANKE data presented in Ref.~\cite{Maeda08}.
However, the effect seems in all cases to be most prominent between
the $K^+K^-$ and $K^0\bar{K}^0$ thresholds. It is therefore natural
to speculate that it is also influenced by virtual $K^0\bar{K}^0$
production and its subsequent conversion into $K^+K^-$ through a
charge-exchange \emph{fsi}. If the $s$-wave $K^+K^-\rightleftharpoons
K^0\bar{K}^0$ coupling is strong, this would generate an observable
cusp at the $K^0\bar{K}^0$ threshold. Such phenomena can
significantly distort spectra as seen, for example, in the case of
the $\Lambda p$ invariant mass distribution from $K^-d\to \Lambda
p\pi^-$ at the $\Sigma N$ threshold~\cite{Tan}.

Cusp effects can be described most economically within the K-matrix
formalism and this, as well as the associated phenomena, has been
discussed extensively by Dalitz and coworkers~\cite{DT,Dalitz}. There
are three basic simplifications that are justified in the treatment
of a problem such as this, where the statistics are low. The first is
that the elements of the K-matrix are taken to be constant,
independent of energy, in the small region from the $K^+K^-$ to a
little above the $K^0\bar{K}^0$ threshold. Secondly, we assume that
isospin invariance is only broken by the mass difference between the
charged and neutral kaons. Finally, since we have only a limited
understanding of the $K\bar{K}$ dynamics, the distortions are taken
to just first order, in which case the resulting formulae have very
transparent interpretations.

\begin{figure}[htb]
\begin{center}
\includegraphics[clip,width=5cm]{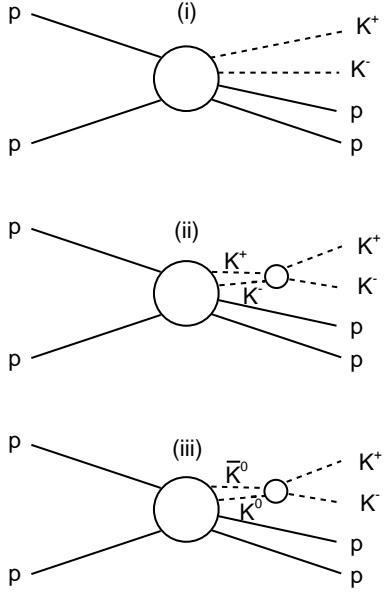}
\caption{Diagrammatic representation of (\emph{i}) direct production
of $K^+K^-$ pairs, where the large blob includes the \emph{fsi}
effects in the $pp$ and $K^-p$ systems considered in
Ref.~\cite{Maeda08}. The elastic $K^+K^-$ \emph{fsi} is illustrated
in (\emph{ii}) and the production of virtual $K^0\bar{K}^0$ pairs
followed by a charge exchange \emph{fsi} in (\emph{iii}).
\label{diagrams}}
\end{center}
\end{figure}

Figure~\ref{diagrams} illustrates the three types of contribution
that are to be considered. In all cases the large blob represents the
$pp\to ppK\bar{K}$ production distorted by the final state
interactions in the $pp$ and $\bar{K}p$ systems, as described in the
simulation presented in Ref.~\cite{Maeda08}. In addition to direct
production of $K^+K^-$ pairs shown in (\emph{i}), these may be
distorted by an elastic rescattering shown in (\emph{ii}). The third
term of panel (\emph{iii}) describes the possibility of the
production of $K^0\bar{K}^0$ pairs that are converted into $K^+K^-$
through a charge exchange \emph{fsi}.

Although the diagrams of Fig.~\ref{diagrams} are easy to visualise in
the charge basis, the evaluation is somewhat simpler in the isospin
basis because the $K\bar{K}$ scattering lengths are normally quoted
in this way. Let $B_0$ and $B_1$ be the \emph{bare} $pp\to
ppK\bar{K}$ amplitudes for producing $s$-wave $K\bar{K}$ pairs in
isospin-0 and 1 states, respectively. These amplitudes, which already
include the \emph{fsi} in the $K^-p$ and $pp$
channels~\cite{Maeda08}, are then distorted through a \emph{fsi}
corresponding to elastic scattering. This leads to enhancement
factors of the form $1/(1-ikA_I)$, where $k$ is the momentum in the
$K^+K^-$ system and $A_I$ is the $s$-wave scattering length in each
of the two isospin channels. The charge-exchange \emph{fsi} of
Fig.~\ref{diagrams}(\emph{iii}) depends upon the $K^0\bar{K}^0\to
K^+K^-$ scattering length, which is proportional to the difference
between $A_0$ and $A_1$, and on the momentum $q$ in the
$K^0\bar{K}^0$ system. In total therefore, the enhancement factor has
a momentum dependence of the form
\begin{equation}
\mathcal{F}=\left|\frac{B_1/(B_1+B_0)}{\left(1-i\frac{1}{2}q[A_1-A_0]\right)(1-ikA_1)}
+\frac{B_0/(B_1+B_0)}{\left(1-i\frac{1}{2}q[A_0-A_1]\right)(1-ikA_0)}\right|^{\,2}\!,
\label{amps}
\end{equation}
where the $\frac{1}{2}$ are isospin factors. In this way we have
extended the $K^-p$ and $pp$ \emph{fsi} factorisation hypothesis of
Ref.~\cite{Maeda08} to include also the $K\bar{K}$ system.

The cusp structure arises because $q$ changes from being purely real
above the $K^0\bar{K}^0$ threshold to purely imaginary below this
point. The strength of the effect clearly depends upon $A_0-A_1$, but
the shape of the signal also depends upon the interference with the
direct $K^+K^-$ production amplitude.

The $K\bar{K}$ scattering lengths are significant due to the presence
of the ($a_0$, $f_0$) scalar resonances~\cite{PDG}, but there is a
large uncertainty in their numerical values, which reflects the
uncertainty in the positions and widths of these states. A useful
summary of the different estimates before 2004 is to be found in
Ref.~\cite{Baru}. It is generally agreed that the isospin $I=1$
scattering length has a small real
part~\cite{Bugg94,Teige99,Achasov03,Antonelli} and we take
$A_1=(0.1+i0.7)\,$fm. Values of the isoscalar scattering length can
be extracted from many fits to
data~\cite{Antonelli,Achasov00,Akhmetshin99,LL} but that deduced by
the BES collaboration, $A_0=(-0.45+i1.63)\,$fm~\cite{BES} seems to be
the most reliable.

There is an even bigger uncertainty in the (complex) ratio of the
production amplitudes,
\begin{equation}\label{Fit_par}
B_1/B_0=C\,\textrm{e}^{i\phi_C}\,,
\end{equation}
which is completely unknown \emph{a priori}. We have therefore fitted
the points shown in Fig.~\ref{Fit} with Eq.~(\ref{amps}) so as to
determine the values of the unknown magnitude $C$ and phase $\phi_C$
within this approach. This has been done separately for the cases
where (\emph{i}) all the terms in Eq.~(\ref{amps}) are retained,
(\emph{ii}) for purely elastic \emph{fsi}, when the $q[A_1-A_0]$
terms are neglected, and (\emph{iii}) purely charge-exchange
\emph{fsi}, when the $kA_I$ terms are discarded. The corresponding
results are presented in Table~\ref{table1} and the associated curves
in Fig.~\ref{Fit}.\\

\begin{center}
\begin{table}[hbt]
\caption{\label{table1}The fit results for the magnitude and phase of
the ratio of the $I=1$ and $I=0$ amplitudes of Eq.~(\ref{Fit_par}).
The data of Fig.~\ref{Fit} are fitted using the \emph{ansatz} of
Eq.~(\ref{amps}) with (i) both elastic and charge-exchange
\emph{fsi}, (ii) elastic \emph{fsi} alone, and (iii) purely
charge-exchange \emph{fsi}. The corresponding curves are shown
together with the experimental points in Fig.~\ref{Fit}.}\vspace{2mm}

\begin{tabular}{|c||c|c|c|}
\hline
Fit par.& el.+c.e.&el.~alone&c.e.~alone\\
\hline
C&$0.62_{-0.11}^{+0.16}$&$0.88_{-0.21}^{+0.05}$&$0.56_{-0.21}^{+0.07}$\\
\hline%
$\phi_C$ (deg)& $-81^{+36}_{-25}$& $159^{+31}_{-4}$&$-131^{+8}_{-20}$\\
\hline
$\chi^2$/\textit{ndf}&1.2&1.2&2.5\\
\hline
\end{tabular}
\end{table}
\end{center}

Although all three sets of curves in Fig.~\ref{Fit} reproduce the
data in an acceptable way, it has to be stressed that the two final
state interactions must be bought as a single package. The effects of
neglecting either the elastic or charge-exchange \emph{fsi} have only
been considered in order to indicate the separate influences of the
two types of contribution.

The full fit of Fig.~\ref{Fit} demonstrates a cusp effect at the
$K^0\bar{K}^0$ threshold but only as a sharp discontinuity in the
slope of the cross section ratio. This is qualitatively similar to
the case where the elastic rescattering is neglected. On the other
hand, the scenario where only elastic rescattering is retained gives
(statistically) an equally good description of the data. This shows a
smooth increase down to the $K^+K^-$ threshold and no anomalous
behaviour at the $K^0\bar{K}^0$ threshold. The nature of the
solutions also differs in that the full one requires mainly $I=0$
production, whereas the contributions from the two isospins would be
rather similar if the channel coupling were disregarded.

\begin{figure}[h]
\begin{center}
\includegraphics[clip,width=7cm, angle=0]{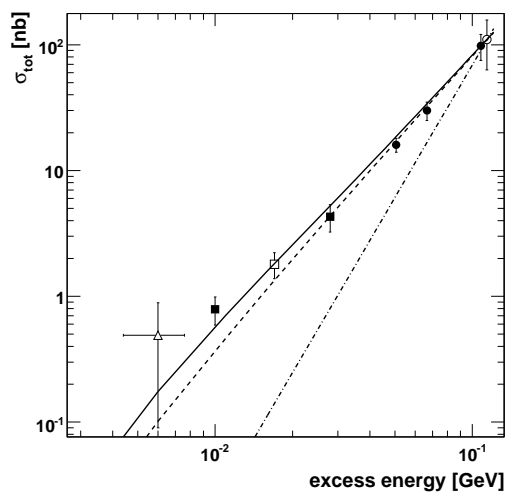}
\caption{Total cross section for the \reactiona\ reaction as a
function of the excess energy. The experimental data are taken from
Refs.~\cite{Maeda08} (closed circles), \cite{DISTO} (open circle),
\cite{Winter} (closed squares), \cite{Wolke} (open triangle), and
\cite{Quentmeier} (open square). The dot-dashed curve is that of
four-body phase space normalised on the 108\,MeV point. The dashed
curve corresponds to the fit which includes final state interactions
between the $K^-$ and the protons and between the two protons
themselves, as described in Ref.~\cite{Maeda08}. The further
inclusion of the \emph{fsi} between the kaon pair leads to the solid
curve. \label{sigt}}
\end{center}
\end{figure}

The $K\bar{K}$ elastic and charge-exchange \emph{fsi} both enhance
the cross section at low masses and, since this region represents a
larger fraction of the total spectrum at low excess energies, it is
clear that these will also affect the energy dependence of the total
production cross section~\cite{Walter}. This is indeed the case, as
shown by Fig.~\ref{sigt}, where the extra contributions from the
\emph{fsi} allow the low and high energy
data to be described simultaneously.\\

The $K\bar{K}$ final state interaction approach used here does not
rely on knowing the basic production mechanism. The analysis of the
\reactiona\ data given in Ref.~\cite{Maeda08} suggests that the main
terms driving the reaction might be linked to $Y^*$ excitation and
decay. However, even if one assumed that the major contribution to
the cross section came from a combination of $a_0$ and $f_0$
production, where these resonances are described by Flatt\'e
shapes~\cite{Flatte}, this would lead to a similar structure to that
of the present work, though only in the vicinity of the $K\bar{K}$
thresholds. Thus the observation of cusps or smooth enhancements at
low $K^+K^-$ invariant mass should not be taken as evidence that the
underlying production mechanism is necessarily driven by the
formation of these scalar resonances.

On the other hand, if $(a_0,f_0)$ production were indeed dominant,
then one could put the kaon mass difference directly into the
Flatt\'e descriptions of these resonances~\cite{CH,Bugg08}. However,
after fitting the \reactiona\ data in terms of $a_0$ and $f_0$
production amplitudes, such a procedure would give results that
differed little from ours in the near-threshold region, provided that
the corresponding values of the scattering lengths were used.

On the basis of the parameters quoted in Table~\ref{table1}, it is
seen that the best fit is achieved with a production of $I=0$
$K\bar{K}$ pairs in the near-threshold region that is about three
times stronger than for $I=1$. This sensitivity originates mainly
from the very different $I=0/I=1$ scattering lengths, which is a
general feature of the various
analyses~\cite{Baru,Bugg94,Teige99,Achasov03,Antonelli,Achasov00,Akhmetshin99,LL,BES}.
This suggests that the production of $I=0$ pairs is dominant in the
\reactiona\ reaction, independent of the exact values of the
scattering lengths.

In summary, on the basis of the existing knowledge of the low energy
$K\bar{K}$ interaction, we would expect there to be a cusp-like
structure in the $K^+K^-$ invariant mass spectrum from the
\reactiona\ reaction. The details, however, are unclear because of
the uncertainties in the relative amplitudes for $I=0$ or $I=1$
production of $K\bar{K}$ pairs, as well as in the $K\bar{K}$
scattering lengths. As seen in Fig.~\ref{Fit}, the data themselves
would be consistent with either a cusp or simply a strong but smooth
low mass enhancement.

Although the energy dependence of the total cross section is better
reproduced when the $K\bar{K}$ rescattering is included, this is
mainly a reflection of it enhancing the cross section at low $K^+K^-$
masses. A similar improvement is found if the charge-exchange
\emph{fsi} is neglected in the fitting process. To establish the
actual nature of the behaviour in the cusp region, better data are
needed and it is hoped that this might be achieved by working
at lower excess energy~\cite{Michael09}.\\

Extensive discussions with Dr.~C.~Hanhart have been particularly
valuable when carrying out the work reported here. We are grateful to
correspondence on this subject with Professors D.V.~Bugg, A.~Gal, and
B.S.~Zou. Support from other members of the ANKE collaboration is
also much appreciated.
%
%

\end{document}